\documentstyle[11pt,cosmos,epsfig]{article}

\begin{document}

\title{Cosmological Perturbation Theory and Structure Formation}

\author{Edmund Bertschinger}
\affil{Department of Physics, MIT Room 6-207, 77 Massachusetts Avenue,
  Cambridge, MA 02139, USA}

\begin{abstract}
These lecture notes discuss several topics in the physics of
cosmic structure formation starting from the evolution of
small-amplitude fluctuations in the radiation-dominated era.
The topics include relativistic cosmological perturbation
theory with the scalar-vector-tensor decomposition,
the evolution of adiabatic and isocurvature initial fluctuations,
microwave background anisotropy, spatial and angular power
spectra, the cold dark matter linear transfer function,
Press-Schechter theory, and a brief introduction to numerical
simulation methods.
\end{abstract}


\section{Introduction}

The advent of precision cosmology with measurements of the Cosmic
Microwave Background (CMB) radiation has brought a new focus on
the physics of the $z>1100$ universe.  The connection between two
fossil relics of the early universe---CMB fluctuations and
galaxies---offers the possibility for improved understanding of
both as well as tighter constraints on cosmological parameters.
These lectures present an overview of the physics of structure
formation starting after the inflationary epoch, going through
recombination, and extending to the present day.

Cosmological perturbation theory is often regarded as a highly
technical subject beyond the scope of graduate courses in cosmology.
However, these lectures may suggest how, with minor simplification,
it can be integrated into a treatment of large scale structure.
The potential benefits are many---a detailed understanding of
CMB anisotropy, a quantitative explanation for the key features
of the cold dark matter transfer function, and an extension of
the intuition developed from the Newtonian theory of large scale
structure to the physics of the post-inflationary universe.
The aim of the current notes is more modest, but it is hoped
that they provide some direction for those wishing to delve
more deeply into cosmological perturbation theory and structure
formation.

\section{Overview of Relativistic Cosmological Perturbation Theory}

The goal of cosmological perturbation theory is to relate the physics
of the early universe (e.g. inflation) to CMB anisotropy and
large-scale structure and to provide the initial conditions for
numerical simulations of structure formation.  The physics during
the period from the end of inflation to the beginning of nonlinear
gravitational collapse is complicated by relativistic effects but
greatly simplified by the small amplitude of perturbations.  Thus,
an essentially complete and accurate treatment of relativistic
perturbation evolution is possible, at least in the context of
simple fluctuation models like inflation.  Numerous articles have
been written on this topic.  See [\cite{lesh,mabert}] for references
and for more detailed treatments similar in style to the present
lecture notes.

The starting point for cosmological perturbation theory is the metric
of a perturbed Robertson-Walker spacetime,
\begin{eqnarray}
  ds^2&=&\left[g_{\mu\nu}^{(0)}+g_{\mu\nu}^{(1)}\right]dx^\mu dx^\nu
    \nonumber\\
      &=&a^2(\tau)\left[-d\tau^2+\gamma_{ij}(\vec x\,)dx^idx^j+
      h_{\mu\nu}(\vec x,\tau)dx^\mu dx^\nu\right]\ .
  \label{rwmet}
\end{eqnarray}
Spatial coordinates take the range $1\le i,j\le 3$; $x^i$ (or
$\vec x$ for all three) is a comoving spatial coordinate; $\tau$
is conformal time; $a(\tau)$ is the cosmic expansion scale factor;
units are chosen so that the speed of light is unity; and
$\gamma_{ij}(\vec x\,)$ is the 3-metric of a maximally symmetric
constant curvature space.  Conformal time is related to the proper
time measured by a comoving observer (i.e. one at fixed $\vec x$)
by $dt=a(\tau)d\tau$.  For cosmological perturbation theory it is
more convenient than proper time.  The metric perturbations are given
by $h_{\mu\nu}=g_{\mu\nu}^{(1)}/a^2$.

The Robertson-Walker model is characterized by one function of time,
$a(\tau)$, and one constant, the curvature constant $K$.  The expansion
factor obeys the Friedmann equation,
\begin{equation}
  \left(\dot a\over a\right)^2={8\pi\over3}Ga^2\bar\rho(a)-K
  \label{friedmann}
\end{equation}
where a dot denotes $d/d\tau$ and $\bar\rho(a)$ is the total mean
energy density.  Note that the usual Hubble parameter is $H=\dot a/a^2$.

The spatial part of the Robertson-Walker metric takes a simple form
in spherical coordinates $(\chi,\theta,\phi)$:
\begin{equation}
  \gamma_{ij}dx^idx^j=d\chi^2+r^2(\chi)\left(d\theta^2+\sin^2\theta
    d\phi^2\right)\ .
  \label{rwdl}
\end{equation}
The angular radius $r(\chi)$ takes one of three forms depending on the
sign of $K$:
\begin{equation}
  r(\chi)=\cases{{1\over\sqrt{K}}\sin(\chi\sqrt{K}),\quad\quad\quad
    K>0,\cr
    \chi,\quad\quad\quad\quad\quad\quad\quad\quad\ \, K=0,\cr
    {1\over\sqrt{-K}}\sinh(\chi\sqrt{-K}),\quad K<0\ .}
  \label{rchi}
\end{equation}
These cases are referred to as closed, flat, and open, respectively.
Many authors choose units of length such that $\chi=\pm1$ or 0.
However, it is convenient to let $K$ have units (inverse length
squared).  From the Friedmann equation, $K=(\Omega_0-1)H_0^2$
where the subscript 0 denotes the present value and $\Omega_0$
is the total density parameter (including dark energy).  Note that
the range of $\chi$ is $[0,\infty)$ in the flat and open models but
is $[0,\pi/\sqrt{K}]$ in the closed model.  The flat Robertson-Walker
model favored by inflation is conformal to Minkowski spacetime,
i.e. the metrics are identical up to an overall conformal factor
$a^2(\tau)$.

\subsection{Scalar-Vector-Tensor Decomposition}

In linear perturbation theory, the metric perturbations $h_{\mu\nu}$
are regarded as a tensor field residing on the background
Robertson-Walker spacetime.  As a symmetric $4\times4$ matrix,
$h_{\mu\nu}$ has 10 degrees of freedom.  Because of the ability
to make continuous deformations of the coordinates, 4 of these
degrees of freedom are gauge- (coordinate-) dependent, leaving
6 physical degrees of freedom.  A proper treatment of cosmological
perturbation theory requires clear separation between physical and
gauge degrees of freedom.

The metric degrees of freedom in linear pertubation theory were
classified by Lifshitz in 1946 [\cite{lifsh}].  Lifshitz presented
the scalar-vector-tensor decomposition of the metric.  It is based
on a $3+1$ split of the components, which we rewrite as follows:
\begin{equation}
  h_{00}\equiv-2\psi\ ,\quad
  h_{0i}\equiv w_i\ ,\quad
  h_{ij}=2\left(\phi\gamma_{ij}+S_{ij}\right)\
    \hbox{with $\gamma^{ij}S_{ij}=0$.}
  \label{3p1}
\end{equation}
Here, $\gamma^{ij}$ is the matrix inverse of $\gamma_{ij}$.
The trace part of $h_{ij}$ has been absorbed into $\phi$ so
that $S_{ij}$ has only 5 independent components.
 
With the space-time ($3+1$) split, we use $\gamma_{ij}$
(or $\gamma^{ij}$) to lower (or raise) indices of spatial 3-vectors
and tensors.  For convenience, spatial derivatives will be
written using the 3-dimensional covariant derivative $\nabla_i$
defined with respect to the 3-metric $\gamma_{ij}$; it is a
three-dimensional version of the full 4-dimensional covariant
derivative presented in general relativity textbooks.  For example,
if $K=0$, we can choose Cartesian coordinates such that $\gamma_{ij}=
\delta_{ij}$ and $\nabla_i=\partial/\partial x^i$.

The scalar-tensor-vector split is based on the decomposition of a vector
into longitudinal and transverse parts.  For any three-vector field
$w_i(\vec x\,)$, we may write
\begin{equation}
  w_i=w_i^\parallel+w_i^\perp\quad\hbox{where}\ \vec\nabla\times
    \vec w^{\,\parallel}=\vec\nabla\cdot\vec w^{\,\perp}=0\ .
  \label{ltsvec}
\end{equation}
The curl and divergence are defined using the spatial covariant
derivative, e.g. $\vec\nabla\cdot\vec w=\gamma^{ij}\nabla_iw_j$.

The longitudinal/transverse decomposition is not unique (e.g. one may
always add a constant to $w_i^\parallel$) but it always exists.  The
terminology arises because in the Fourier domain $w_i^\parallel$ is
parallel to the wavevector while $w_i^\perp$ is transverse (perpendicular
to the wavevector).  Note that $w_i^\parallel=\nabla_i\phi_w$ for some
scalar field $\phi_w$.  Thus, the longitudinal/transverse decomposition
allows us to write a vector field in terms of a scalar (the longitudinal
or irrotational part) and a part that cannot be obtained from a scalar
(the transverse or rotational part).

A similar decomposition holds for a two-index tensor, but now each index
can be either longitudinal or transverse.  For a symmetric tensor, there
are three possibilities: both indices are longitudinal, one is transverse,
or two are transverse.  These are written as follows:
\begin{equation}
  S_{ij}=S_{ij}^\parallel+S_{ij}^\perp+S_{ij}^{\rm T}\ ,
  \label{ltsten}
\end{equation}
where
\begin{equation}
  \gamma^{jk}\nabla_kS_{ij}=\gamma^{jk}\nabla_kS_{ij}^\parallel+
    \gamma^{jk}\nabla_kS_{ij}^\perp\ .
  \label{divtens}
\end{equation}
The first term in equation (\ref{divtens}) is a longitudinal vector
while the second term is a transverse vector.  The divergence of the
doubly-transverse part, $S_{ij}^{\rm T}$, is zero.  For a traceless
symmetric tensor, the doubly and singly longitudinal parts can be
obtained from the gradients of a scalar and a transverse vector,
respectively:
\begin{equation}
  S_{ij}^\parallel=\left(\nabla_i\nabla_j-{1\over3}\gamma_{ij}
    \nabla^2\right)\phi_S\ ,\quad
  S_{ij}^\perp=\nabla_iS_j^\perp+\nabla_jS_i^\perp\ .
  \label{longtens}
\end{equation}

Although we will not prove it here (see [\cite{lesh}] for the details),
under an infinitesimal coordinate transformation $S_{ij}^\parallel$
and $S_{ij}^\perp$ can change while $S_{ij}^{\rm T}$ is invariant.
Similarly, the longitudinal and transverse parts of $w_i=h_{0i}$
are both gauge-dependent.

Now we have the mathematical background needed to perform the
scalar-tensor-vector split of the physical degrees of freedom
of the metric.  The ``tensor mode'' represents the part of $h_{ij}$
that cannot be obtained from the gradients of a scalar or
vector, namely $S_{ij}^{\rm T}$.  The tensor mode is gauge-invariant
and has two degrees of freedom (five for a symmetric traceless
$3\times3$ matrix, less three from the condition $\gamma^{jk}
\nabla_kS_{ij}^{\rm T}=0$).  Physically it represents gravitational
radiation; the two degrees of freedom correspond to the two
polarizations of gravitational radiation.  Gravitational radiation
is transverse: a wave propagating in the $z$-direction can have
nonzero components $h_{xx}-h_{yy}$ and $h_{xy}=h_{yx}$ but no others.
The tensor mode $S_{ij}^{\rm T}$ behaves like a spin-2 field under
spatial rotations.  Note that my definition of the tensor mode
strain $S_{ij}$ here differs by a factor of 2 from the conventional
$h_{ij}^{TT}$ in [\cite{mtw}].

The ``vector mode'' behaves like a spin-1 field under spatial
rotations.  It corresponds to the transverse vector parts of the
metric, which are found in $w_i^\perp$ and $S_{ij}^\perp$.  Each
part has two degrees of freedom.  Although we will not prove it
here (see [\cite{lesh}] for the details), it is possible to eliminate
two of these degrees of freedom by imposing gauge conditions
(coordinate conditions).  The synchronous gauge of Lifshitz
[\cite{lifsh}] is one popular choice, with $w_i^\perp=w_i^\parallel=0$.
Another choice is the ``Poisson'' or transverse gauge [\cite{lesh}]
which sets $S_{ij}^\perp=0$.  In either case, there are two
physical degrees of freedom and they correspond physically
to gravitomagnetism.  Although this effect is less well known than
gravitational radiation or Newtonian gravity, it produces magnetic-like
effects on moving and spinning masses, such as the precession of a
gyroscope in the gravitational field of a spinning mass (Lense-Thirring
precession).  This phenomenon has not yet been discovered experimentally
but should be measured by the Gravity Probe B satellite to be launched
in 2002 [\cite{gpb}].

The ``scalar mode'' is spin-0 under spatial rotations and corresponds
physically to Newtonian gravitation with relativistic modifications.
The scalar parts of the metric are given by $\phi$, $\psi$,
$w_i^\parallel$, and $S_{ij}^\parallel$.  Any two of these may
be set to zero by means of a gauge transformation.  One popular choice
is $w_i^\parallel=S_{ij}^\parallel=0$, also known as the conformal
Newtonian gauge.  It corresponds to the scalar mode in the {\bf transverse
gauge}, defined by the gauge conditions
\begin{equation}
  \gamma^{ij}\nabla_iw_j=0\ ,\quad
  \gamma^{jk}\nabla_kS_{ij}=0\ .
  \label{transg}
\end{equation}
This gauge in linearized general relativity is the gravitational
analogue of Coulomb gauge in electromagnetism, where the magnetic
vector potential is transverse, $\nabla_iA^i=0$.  In the gravitational
transverse gauge, $w_i$ is transverse and $S_{ij}$ is doubly transverse.
This gauge is convenient for developing intuition although not necessarily
the best for computation.  The variables $(\phi,\psi,w_i^\perp,
S_{ij}^{\rm T})$ correspond to gauge-invariant variables introduced
by Bardeen [\cite{bard}] as linear combinations of metric variables
in other gauges.

In linear perturbation theory, the scalar, vector, and tensor modes
evolve independently.  The vector and tensor modes produce no density
perturbations and therefore are unimportant for structure formation,
although they do perturb the microwave background.  In the remainder
of these lectures, only the scalar mode will be considered.

The Einstein equations give the equations of motion for the metric
perturbations in terms of the energy-momentum tensor, the source of
relativistic gravity.  Here we consider the case of a perfect fluid
(or several perfect fluid components combined), for which
\begin{equation}
  T^{\mu\nu}=(\rho+p)V^\mu V^\nu+pg^{\mu\nu}
  \label{tmunu}
\end{equation}
where $\rho$ and $p$ are the proper energy density and pressure in
the fluid rest frame and $V^\mu$ is the fluid 4-velocity.

We split $T^{\mu\nu}$ into time and space components as we did for the
metric.  It is convenient to use mixed components because the conformal
factor $a^2$ then cancels, yielding
\begin{eqnarray}
  \label{tcompon}
  T^0_{\ \,0}&=&-\rho(\vec x,\tau)=-\left[\bar\rho(\tau)+\delta
    \rho(\vec x,\tau)\right]\ ,\nonumber \\
  T^0_{\ \,i}&=&\left[\bar\rho(\tau)+\bar p(\tau)\right]v_i(\vec x,\tau)
    =-\left[\bar\rho(\tau)+\bar p(\tau)\right]\nabla_i W\ ,\\
  T^i_{\ \,j}&=&\left[\bar p(\tau)+\delta p(\vec x,\tau)\right]
    \delta^i_{\ \,j}\ ,\nonumber
\end{eqnarray}
where $\bar\rho(\tau)$ and $\bar p(\tau)$ are respectively the energy
(or mass) density and pressure of the Robertson-Walker background
spacetime, $v_i=\gamma_{ij}dx^j/d\tau$ is the fluid 3-velocity
(assumed nonrelativistic), and $W$ is a velocity potential.  We are
assuming that the velocity field is irrotational or, if it is not,
that the matter fluctuations are linear so that the longitudinal
and transverse velocity components evolve independently.  This is
a good approximation prior to the epoch of galaxy formation.

We are also making another approximation, namely that the shear
stress (the non-diagonal part of $T^i_{\ \,j}$) is negligible
compared with the pressure.  This is a bad approximation for
relativistic neutrinos, whose free-streaming leads to a non-isotropic
momentum distribution after neutrino decoupling at a temperature
of about 1 MeV.  (A similar but smaller effect occurs for photons
after they decouple at a temperature of about 0.3 eV.)  However,
the effect of this error is modest during the radiation-dominated
era (photons contribute more pressure than neutrinos, and are
isotropic) and is unimportant during the matter-dominated era,
when the nonrelativistic matter density is the major source of
gravity.  We introduce this approximation here in order to
simplify the treatment; full calculations include the effects of
neutrino shear stress [\cite{mabert}].

When the stress tensor is isotropic, the Einstein equations in
transverse gauge imply that the two scalar potentials $\phi$ and
$\psi$ are equal [\cite{lesh}] and the metric takes the simple form
\begin{equation}
  ds^2=a^2(\tau)\left[-(1+2\phi)d\tau^2+(1-2\phi)\gamma_{ij}dx^idx^j
    \right]\ .
  \label{cnmet}
\end{equation}
This is a simple cosmological generalization of the standard weak-field
limit of general relativity, which is recovered by setting $a=1$ and
$\gamma_{ij}=\delta_{ij}$.

Given the stress-energy tensor, we can now obtain field equations for
$\phi(\vec x,\tau)$ using the Einstein equations.  The only independent
equation for the Robertson-Walker background is the Friedmann equation
(\ref{friedmann}).  For the perturbations, we obtain
\begin{equation}
  \label{poisson}
  (\nabla^2+3K)\phi=4\pi Ga^2\left[\delta\rho+3{\dot a\over a}(\bar\rho
    +\bar p)W\right]\ ,
\end{equation}
\begin{equation}
  \label{momcon}
  \partial_\tau\phi+{\dot a\over a}\phi=4\pi Ga^2(\bar\rho+\bar p)W\ ,
\end{equation}
and
\begin{equation}
  \label{phidd}
  \partial_\tau^2\phi+3{\dot a\over a}\partial_\tau\phi-\left(
    8\pi Ga^2\bar p+2K\right)\phi=4\pi Ga^2\delta p\ .
\end{equation}
There are more equations than in Newtonian gravity because the
Einstein equations have local energy-momentum conservation built
into them.

Equation (\ref{poisson}) is remarkably like the Poisson equation of
Newtonian gravity, especially when one realizes that the factor of
$a^2$ is needed to convert the Laplacian from comoving to proper
spatial coordinates.  Otherwise there are three differences.  First,
the spatial curvature adds a scale to the Laplacian in a curved space.
Second,  the density source is $\delta\rho=\rho-\bar\rho$ rather
than $\rho$ because $\phi$ represents a metric perturbation on top
of the background Robertson-Walker spacetime.  Third, momentum density
(through $W$) is a source for gravity in general relativity.
However, this term is comparable with the density term only on
length scales comparable to or larger than the Hubble length $H^{-1}$.

Two of our field equations are unfamiliar from Newtonian dynamics.
Equation (\ref{momcon}) is reminiscent of the Zel'dovich approximation
[\cite{za}] in that the velocity potential is proportional to the
gravitational potential in the linear regime when $\phi(\vec x,\tau)=
\phi_0(\vec x\,)D_\phi(\tau)$.  However, it is a more general result
whose Newtonian limit may be derived by combining the continuity
equation with the time derivative of the Poisson equation.

Equation (\ref{phidd}) is more surprising, as it gives the
gravitational potential through a second-order time evolution
equation as opposed to the action-at-a-distance Poisson equation
(\ref{poisson}).  The reader may well wonder about the compatibility
of the two equations as well as the physical validity of the Poisson
equation in general relativity, but equations (\ref{poisson})--(\ref
{phidd}) are all correct.  See [\cite{lesh}] for a full discussion.
In brief, causality is restored by the tensor mode.  The time
evolution equation for the potential can be regarded as arising
from local energy-momentum conservation (which is built into the
Einstein equations) combined with the Poisson equation.  Indeed,
an equation similar to equation (\ref{phidd}) can be derived in
Newtonian gravity from the fluid and Poisson equations.  Its utility
arises from the fact that when pressure perturbations are
gravitationally unimportant (i.e. when matter and/or dark energy
dominate over radiation), the time evolution of perturbations can
be obtained from a single differential equation depending only on
the background curvature, pressure, and density.

\subsection{Perfect Fluid Model}

To conclude this section, we present a simple model for understanding
CMB anisotropy and large scale structure.  We reduce the matter and
energy content of the universe to a vacuum energy (with $\rho_v=
\Lambda/8\pi G$ contributing to the mean expansion rate but not to
the fluctuations) and two fluctuating components, CDM and radiation
(photons and neutrinos, the former coupled to baryons by electron
scattering until recombination).  The CDM and radiation are treated
as perfect fluids, ignoring the free-streaming of neutrinos and the
diffusion and streaming of photons during and after recombination.

The fluid equations follow from local energy-momentum conservation,
$\nabla_\mu T^{\mu\nu}=0$.  For a perfect fluid, with the scalar mode
only (no vorticity and no gravitational radiation), the fluid equations
in the metric of equation (\ref{cnmet}) are
\begin{equation}
  \label{conteq}
  \partial_\tau\rho+3\left({\dot a\over a}-\partial_\tau\phi\right)
  (\rho+p)+\nabla_i\left[(\rho+p)v^i\right]=0
\end{equation}
and
\begin{equation}
  \label{eulereq}
  \partial_\tau\left[(\rho+p)v^i\right]+4{\dot a\over a}(\rho+p)v^i
    +\nabla_ip+(\rho+p)\nabla_i\phi=0\ .
\end{equation}
These are familiar from Newtonian fluid dynamics, with some extra terms.
The damping ($\dot a/a$) terms arise from Hubble expansion and the use
of comoving coordinates.  The $\partial_\tau\phi$ term in the continuity
equation (\ref{conteq}) arises from the deformation of the spatial
coordinates, i.e. the $\phi\delta_{ij}$ contribution to the metric.
(The flux term in eq. \ref{conteq} describes transport relative to the
comoving coordinate grid, which is deforming when $\partial_\tau\phi\ne0$.)
The energy flux (or momentum density, they are equal) includes pressure
in relativity because of the $pdV$ work done by compression.
Equations (\ref{conteq}) and (\ref{eulereq}) apply separately to
perfect fluid component.

\section{Adiabatic and Isocurvature Modes}

This section discusses the evolution of perturbations from the end
of inflation (taken to be at $\tau=0$ for all practical purposes)
through recombination using a two-fluid approximation.  We approximate
the matter and energy content in the universe as being two perfect
fluids: Cold Dark Matter (a zero-temperature gas) and radiation
(photons coupled to electrons and baryons before recombination,
plus neutrinos which are approximated as behaving like photons).
The net energy density perturbation is thus
\begin{equation}
  \label{delrho}
  \delta\rho(\vec x,\tau)=\bar\rho_c(a)\delta_c(\vec x,\tau)+
    \bar\rho_r(a)\delta_r(\vec x,\tau)\ ,
\end{equation}
where $\bar\rho_c\propto a^{-3}$ and, if we neglect the contribution
of baryons, $\bar\rho_r\propto a^{-4}$.  We could add a third
component for the baryons, but this complicates the presentation
without adding essential new behavior.  In the interest of economy,
we leave baryons out.  We define the relative density contrasts
$\delta_c\equiv\delta\rho_c/\bar\rho_c$ and $\delta_r\equiv
\delta\rho_r/\bar\rho_r$.

The total pressure is simply $p(\vec x,\tau)={1\over3}\bar\rho_r(a)
(1+\delta_r)+p_v$ where $p_v=-\rho_v$ is the spatially constant
negative pressure of vacuum energy (cosmological constant), should
any be present.

As we noted in the previous section, density perturbations couple
gravitationally only to the scalar mode of metric perturbations,
and the scalar mode cannot generate transverse vector fields.
As a result, the peculiar velocity fields of CDM and radiation are
longitudinal and are fully characterized by their potentials
$W_c$ and $W_r$ for CDM and radiation, respectively.

As a final simplifying assumption, we will assume that we are
studying effects on distance scales less than the curvature
length $\vert K\vert^{-1/2}=\vert\Omega_0-1\vert^{-1/2}H_0^{-1}$,
which is at least 5 Gpc and possibly infinitely large.

With these assumptions, we can linearize the fluid equations
(\ref{conteq}) and (\ref{eulereq}) for $\delta_c^2\ll1$ and
$\delta_r^2\ll1$ to obtain
\begin{eqnarray}
  \label{linfluid}
  \partial_\tau\delta_c=\nabla^2W_c+3\partial_\tau\phi\ ,&&
    \partial_\tau\delta_r={4\over3}\nabla^2W_r+4\partial_\tau\phi\ ,
    \nonumber \\
  \partial_\tau W_c+{\dot a\over a}W_c=\phi\ ,&&
    \partial_\tau W_r={1\over4}\delta_r+\phi\ .
\end{eqnarray}
These equations are easy to understand.  The $\partial_\tau\phi$
terms have the same origin as in equation (\ref{conteq}),
namely the deformation of the comoving coordinate system.
The velocity potential terms are related to the velocity
divergence, $\vec\nabla\cdot\vec v=-\nabla^2 W$.
The radiation density fluctuations grow at $4/3$ the rate
of matter fluctuations to compensate for the fact that
the photon number density is proportional to $\rho_r^{3/4}$.
As expected, the velocity equations show that matter
velocities redshift (for $\phi=0$) with $\vec v_c\propto
a^{-1}$ while radiation velocities (e.g. the speed of light)
do not redshift.  (Momenta redshift $\propto a^{-1}$ but
the speed of light is unaffected by expansion.)  Finally,
radiation pressure gives rise to an acoustic restoring force
$-\vec\nabla p_r=-(1/3)\bar\rho_r\vec\nabla\delta_r$ which
shows up as the $(1/4)\delta_r$ term in the radiation
longitudinal velocity equation.

Equations (\ref{linfluid}) are coupled by gravity, which can
be determined from the Poisson equation (\ref{poisson}), which
becomes
\begin{equation}
  \label{poisson2}
  \nabla^2\phi=4\pi Ga^2\left[(\bar\rho_r\delta_r+\bar\rho_c
  \delta_c)+{\dot a\over a}(4\bar\rho_rW_r+3\bar\rho_cW_c)
    \right]\ .
\end{equation}

We can solve equations (\ref{linfluid}) and (\ref{poisson2})
most easily by expanding the spatial dependence in eigenfunctions
of the Laplacian $\nabla^2$.  In flat space these are plane
waves $\exp(i\vec k\cdot\vec x\,)$, and this is a good approximation
even if the background is curved, provided $k\gg\vert K\vert^{1/2}$.
This approximation is valid for almost all applications except
large angular scale CMB anisotropy.  Note that $\vec k$ is a
comoving wavevector; the physical wavelength is $2\pi a/k$.

By expanding the spatial dependence in plane waves, we now have
a fourth-order linear system of ordinary differential equations
in time.  We are free to define linear combinations of the
variables $(\delta_c,\delta_r,W_c,W_r)$ as our fundamental
variables.  In terms of the mechanisms for generating primeval
perturbations, the most natural variables are the metric
perturbation $\phi$ and the specific entropy
\begin{equation}
  \label{eta}
  \eta\equiv{\delta p-c_{\rm s}^2\delta\rho\over\bar\rho_c
    c_{\rm s}^2}={3\over4}\delta_r-\delta_c\ ,
\end{equation}
where
\begin{equation}
  \label{cs2}
  c_{\rm s}^2\equiv{d\bar p/d\tau\over d\bar\rho/d\tau}=
    \left[3\left(1+{3\bar\rho_c\over4\bar\rho_r}\right)\right]^{-1}
\end{equation}
is the effective one-fluid sound speed of the matter and
radiation.  (Acoustic signals actually do not propagate at
this speed; they propagate with speed $3^{-1/2}$ through
the radiation and they do not propagate at all through CDM.)

For a two-component radiation plus CDM universe, the
solution of the Friedmann equation (\ref{friedmann}) gives
\begin{equation}
  \label{ytau}
  y\equiv{\bar\rho_c\over\bar\rho_r}={a(\tau)\over a_{\rm eq}}
    ={\tau\over\tau_e}+\left(\tau\over2\tau_e\right)^2\ ,\ \
  \tau_e\equiv\left(a_{\rm eq}\over\Omega_c\right)^{1/2}
    H_0^{-1}={19\,{\rm Mpc}\over\Omega_ch^2}\ .
\end{equation}
The radiation-dominated era ends and matter-dominated era
begins at redshift $1+z_{\rm eq}=a^{-1}_{\rm eq}=2.5\times
10^4\Omega_mh^2$.  A cosmological constant has no significant
effect on $y(\tau)$ provided that, during the times of
interest, the vacuum energy density is much less than the
radiation or cold dark matter density.

The fluid and Poisson equations can now be combined to
give a pair of second-order ordinary differential equations,
\begin{eqnarray}
  \label{phietaeqs}
  {1\over3c_{\rm s}^2}\partial_\tau^2\phi+\left(1+{1\over
    c_{\rm s}^2}\right){\dot a\over a}\partial_\tau\phi+
    \left({k^2\over3}+{1\over4y\tau_e^2}\right)\phi
    &=&{\eta\over2y\tau_e^2}\ ,\nonumber \\
  {1\over3c_{\rm s}^2}\partial_\tau^2\eta+{\dot a\over a}
    \partial_\tau\eta+{k^2y\over4}\eta&=&{1\over6}
    y^2k^4\tau_e^2\phi\ .
\end{eqnarray}
It is interesting to note that $\phi$ and $\eta$ evolve
independently aside from the source term each provides
to the other.  The coupling implies that entropy
perturbations are a source for the growth of gravitational
perturbations and vice versa.

For a given wavenumber $k$, there are two key times in the
evolution of $\phi$ and $\eta$: the sound-crossing time
$\tau\approx \pi/(kc_{\rm s})$ and the time of matter-radiation
equality at $\tau\approx\tau_e$.  [To be precise, $a=a_{\rm eq}$
at $\tau/\tau_e=2(\sqrt{2}-1)$.]  For $\tau\ll\tau_{\rm e}$,
$\phi$ and $\eta$ decouple and they both have solutions
that decay rapidly with $\tau$ as well as solutions that
are finite as $\tau\to0$.  The latter are conventionally
called ``growing modes'' while the former are called decaying.
The growing mode solution in the radiation era $\tau\ll\tau_e$ is
\begin{eqnarray}
  \label{phietagrow}
  \phi(\vec k,\tau)&=&{3\over(\omega\tau)^3}(\sin\omega\tau
    -\omega\tau\cos\omega\tau)A(\vec k\,)\nonumber \\
    &&+{\tau\over\tau_e}\left[{1\over(\omega\tau)^4}+{1\over
    2(\omega\tau)^2}-{1\over(\omega\tau)^4}(\cos\omega\tau
    +\omega\tau\sin\omega\tau)\right]I(\vec k\,)\ , \\
  \eta(\vec k,\tau)&=&I(\vec k\,)+9\left[\ln\omega\tau
    +{\cal C}-{\rm Ci}(\omega\tau)+{1\over2}(\cos\omega\tau-1)
    \right]A(\vec k\,)\ ,\nonumber
\end{eqnarray}
where $\omega\equiv k/\sqrt{3}$ is the phase of acoustic waves
in the radiation fluid, ${\cal C}=0.5772\ldots$ is the Euler
constant, and ${\rm Ci}(x)$ is the cosine integral defined by
\begin{equation}
  \label{cosint}
  {\rm Ci}(x)={\cal C}+\ln x+\int_0^x{\cos t-1\over t}\,dt\ .
\end{equation}
In equation (\ref{phietagrow}) we have neglected all terms
$O(\tau/\tau_e)$ aside from the lowest-order contribution of
$I(\vec k\,)$ to $\phi$.

The general growing-mode solution contains two $k$-dependent
integration constants, $A(\vec k\,)$ and $I(\vec k\,)$, which
represent the initial ($\tau\to0$) amplitude of $\phi$ and
$\eta$, respectively.  Conventionally they are called the
``adiabatic'' (so-called because the entropy perturbation
$\eta=0$ at $\tau=0$, although ``isentropic'' would be more
appropriate) and ``isocurvature'' (so-called because
$\phi=0$ at $\tau=0$) mode amplitudes.  In addition, there
are decaying modes whose amplitude at $\tau=\tau_e$ is
negligible unless the initial conditions were fine-tuned
or there are external sources such as topological defects
to continually regenerate the decaying mode.  The solution
of equations (\ref{phietagrow}) is valid only for inflation
or other mechanisms that generate fluctuations exclusively
in the early universe, but not for topological defects.

As a result of reheating, inflation produces fluctuations
with spatially constant ratios of the number densities of
all particle species.  For CDM and radiation, for example,
$\rho_r^{3/4}/\rho_c=\hbox{constant}$ hence $I=0$ and thus
$\eta=0$ until radiation pressure forces begin to separate
the radiation from the CDM, which takes a sound-crossing
time or $\tau\sim\omega^{-1}$.  Thus, although inflation produces
the adiabatic mode of initial fluctuations, the differing equations
of state $p(\rho)$ for the two-fluid system lead to entropy
perturbations proportional to the initial adiabatic mode
amplitude $A(\vec k\,)$ (see the second of eqs. \ref{phietagrow}).

Physically, the number density of particles of all species
are proportional to each other in the adiabatic mode as long
as there has been too little time for particles to travel a
significant difference via their thermal speeds.  However,
the differing thermal speeds of different particle species
(e.g. photons and cold dark matter) eventually cause the
particle densities to evolve differently, so that the
number density ratios are no longer constant and the entropy
perturbation is no longer zero.

First-order phase transitions, on the other hand, change
the equation of state without moving matter or energy on
macroscopic scales, hence $\delta\rho_r+\delta\rho_c=
\theta_c=\theta_r=0$ initially implying $\phi=0$ initially
but $\eta\ne0$.  This isocurvature mode is characterized
by the initial entropy perturbation amplitude $I(\vec k\,)$.
As the first of equations (\ref{phietagrow}) shows, the
isocurvature mode generates curvature fluctuations when
the universe becomes matter-dominated.

The isocurvature CDM inflationary model has a fine balance
such that $\delta_r+y\delta_c=0$ initially.  This may involve
nonlinear fluctuations of $\delta_c$ at early times, but
$\delta_r$ is negligible early in the radiation-dominated
era $y\ll1$.  For this reason, the isocurvature mode was
originally called isothermal [\cite{pd68}].

On small scales, $k\tau\gg1$, equations (\ref{phietaeqs})
may be solved using a WKB approximation.  For our purposes
it suffices to take the $\omega\tau\gg1$ limit of equations
(\ref{phietagrow}):
\begin{equation}
  \label{phisubhor}
  \phi\approx-3{\cos\omega\tau\over(\omega\tau)^2}A(\vec k\,)
    +{1\over\omega^3\tau_e\tau^2}\left({\omega\tau\over2}-\sin
    \omega\tau\right)I(\vec k\,)\ .
\end{equation}
The radiation density fluctuation scales as
\begin{displaymath}
  \delta_r\sim{k^2\phi\over Ga^2\bar\rho_r}\sim k^2\tau^2
    \phi\ ,
\end{displaymath}
which oscillates in proportion to $cos\omega\tau$ for the
adiabatic mode and grows in proportion to $y\sim\tau/\tau_e$
for the isocurvature mode.

Note from equation (\ref{phisubhor}) that the isocurvature
mode induces oscillations in the potential with a $\pi/2$
phase shift relative to the adiabatic mode.  This shift
translates into a shift in the positions of the acoustic
peaks in the CMB angular power spectrum.

\subsection{The Adbiabatic Growing Mode as a Standing Wave}

It is instructive to consider the general adiabatic perturbation
in the radiation-dominated era including the decaying mode.
This problem is easy to analyze and shows the role of
acoustic waves in the radiation-dominated era.

The two linearly independent solutions of the first of
equations (\ref{phietaeqs}) for $y\ll1$ and adiabatic
initial conditions ($\eta=0$) are
\begin{equation}
  \label{phigd}
  \phi_g={3\over(\omega\tau)^3}(\sin\omega\tau-\omega\tau
    \cos\omega\tau)\ ,\ \
  \phi_d={3\over(\omega\tau)^3}(\cos\omega\tau+\omega\tau
    \sin\omega\tau)
\end{equation}
where the subscripts label the growing and decaying modes,
respectively.  They are the real and imaginary parts of
\begin{equation}
  \label{phiomega}
  \phi_\omega=\phi_g+i\phi_d={3\over(\omega\tau)^3}
    (i-\omega\tau)e^{-i\omega\tau}\ .
\end{equation}

For a given spatial harmonic, $\exp(i\vec k\cdot\vec x\,)
\phi_\omega$ represents a wave traveling in the $\vec
k$-direction.  Thus, the growing mode $\phi_g={1\over2}
(\phi_\omega+\phi_\omega^\ast)$ represents a superposition
of waves traveling along $\vec k$ and $-\vec k$, i.e. a
standing wave
\begin{equation}
  \label{standing}
  \phi_g\cos kx={\cal R}e\ {3\over2(\omega\tau)^3}\left[
    (i-\omega\tau)e^{i(\vec k\cdot\vec x-\omega\tau)}
    +(-i-\omega\tau)e^{i(\vec k\cdot\vec x+\omega\tau)}
    \right]\ .
\end{equation}
The waves travel at the sound speed $\omega/k=1/\sqrt{3}$
in the radiation fluid.

The entropy perturbation induced by a traveling wave potential
perturbation is easily calculated in the radiation-dominated
era using the Green's function method applied to the
second of equations (\ref{phietaeqs}) with $\phi=\phi_\omega$
for the source term.  The result is
\begin{equation}
  \label{etaomega}
  \eta_\omega=9\int_0^{\omega\tau}{1\over x}\left(1-e^{-ix}
    \right)dx-{9\over2}\left(1-e^{-i\omega\tau}\right)\ .
\end{equation}
The real part of this equation gives the $A(\vec k\,)$
contribution to $\eta$ in equations (\ref{phietagrow}).

\section{CMB Anisotropy}

This section presents a simplified treatment of CMB anisotropy,
with the aim of highlighting the essential physics without
getting lost in all of the details.  More complete treatments are
found in [\cite{cmbfast,hswz98}].

The microwave background radiation is fully described by
a set of photon phase space distribution functions.
Ignoring polarization (a few percent effect), all the
information is included in the intensity  or in $f(\vec x,
\vec p,\tau)$ where $fd^3p$ is the number density of photons
of conjugate momentum $\vec p$ at position and time
($\vec x,\tau)$.  The conjugate momentum is related
to the proper momentum measured by a comoving observer,
$\vec p/a$, so that $\vec p=\hbox{constant}$ along a
photon trajectory in the absence of metric perturbations.

Remarkably, despite metric perturbations and scattering
with free electrons, the CMB photon phase space distribution
remains blackbody (Planckian) to exquisite precision:
\begin{equation}
  \label{fbb}
  f(\vec x,\vec p,\tau)=f_{\rm Planck}\left(E\over kT
    \right)=f_{\rm Planck}\left[{p\over kT_0(1+\Delta)}\right]
\end{equation}
where $T_0=2.728$ K is the current CMB temperature and
$\Delta(\vec x,\vec n,\tau)$ is the temperature variation
at position $(\vec x,\tau)$ for photons traveling in
direction $\vec n$.  The phase space density is blackbody
but the temperature depends on photon direction as a
result of scattering and gravitational processes occurring
along the line of sight.

The phase space density may be calculated from initial conditions
in the early universe through the Boltzmann equation
\begin{equation}
  \label{boltzmann}
  {Df\over D\tau}\equiv{\partial f\over\partial\tau}+{\partial f
    \over\partial\vec x}\cdot{d\vec x\over d\tau}+{\partial f
    \over\partial\vec p}\cdot{d\vec p\over d\tau}=\left(
    df\over d\tau\right)_c
\end{equation}
where the right-hand side is a collision integral coming from
nonrelativistic electron-photon elastic scattering.  In the
absence of scattering, the phase space density is conserved
along the trajectories
\begin{equation}
  \label{geodes}
  {d\vec x\over d\tau}=\vec n\equiv{\vec p\over p}\ ,\ \
    {1\over p}{dp\over d\tau}=-\vec n\cdot\vec\nabla\phi
    +\partial_\tau\phi\ ,\ \
    {d\vec n\over d\tau}=-2\left[\vec\nabla-\vec n\,(\vec n
    \cdot\vec\nabla)\right]\phi\ .
\end{equation}
Note that the photon energy changes via gravitational redshift
(the first term in $dp/d\tau$) or a time-changing potential,
but in either case the energy of all photons is changed by
the same factor.  The photon direction changes by transverse
deflections due to gravity; $d\vec n/d\tau$ is twice the
Newtonian value, a result well-known in gravitational lens
theory [\cite{gl}].  Because $\partial f/\partial\vec n=0$
for the unperturbed background (the anisotropy arises due to
the metric perturbations), gravitational lensing affects the
CMB anisotropy only through nonlinear effects on small scales
[\cite{selgl}].

The procedure for computing CMB anisotropy is to linearize
the Boltzmann equation assuming $\Delta^2\ll1$ [\cite{peeyu}].
Until recently, the temperature anisotropy $\Delta$ was
expanded in spherical harmonics and the Boltzmann equation
was solved as a hierarchy of coupled equations for the
various angular moments [\cite{mabert}].  In 1996, Seljak
and Zaldarriaga [\cite{cmbfast}] developed a much faster
integration method call CMBFAST based on integrating the
linearized Boltzmann equation along the observer's line of
sight before the angular expansion is made:
\begin{equation}
  \label{cmblos}
  \Delta(\vec n,\tau_0)=\int_0^{\tau_0}d\chi\,e^{-\tau_{\rm T}
    (\chi)}\left[-\vec n\cdot\vec\nabla\phi+\partial_\tau\phi
    +an_e\sigma_{\rm T}\left({1\over4}\delta_\gamma+
    \vec v_e\cdot\vec n+\hbox{pol. term}\right)\right]_
    {\rm ret}
\end{equation}
where $\tau_0$ is the present conformal time, $\chi$ is the
radial comoving coordinate of equation (\ref{rwdl}),
subscript ``ret'' denotes evaluation using retarded time
$\tau=\tau_0-\chi$, $\delta_\gamma$ is the relative
density fluctuation in the photon gas ($\delta_\gamma=
\delta_r$ in our two-fluid approximation), $\vec v_e$
is the mean electron velocity (i.e. the baryon velocity,
which equals $\vec v_r$ in our two-fluid approximation),
$\sigma_{\rm T}$ is the Thomson cross section, and the
Thomson optical depth is
\begin{equation}
  \label{thomsopt}
  \tau_{\rm T}(\chi)\equiv\int_0^\chi d\chi\,(an_e
    \sigma_{\rm T})_{\rm ret}\ .
\end{equation}
We have left out small terms (``pol. term'') due to
polarization and the anisotropy of Thomson scattering in
equation (\ref{cmblos}).

The terms in equation (\ref{cmblos}) are easy to understand.
The $\exp(-\tau_{\rm T})$ factor accounts for the opacity
of electron scattering, which prevents us from seeing much
beyond a redshift of 1100, where $\tau_{\rm T}\approx1$.
The CMB anisotropy appears to come from a thin layer called
the photosphere, just like the radiation from the surface
of a star.

The two gravity terms give the effective emissivity due
to the photon energy change caused by a varying gravitational
potential.  For a blackbody distribution, if all photon
energies are increased by a factor $1+\epsilon$, the
distribution remains blackbody with a temperature increased
by the same factor.  Thus, the line-of-sight integration
of the fractional energy change of equation (\ref{geodes})
translates directly into a temperature variation.

The terms proportional to the Thomson opacity $an_e
\sigma_{\rm T}$ are the effective emissivity due to
Thomson scattering.  Bearing in mind that the photon-baryon
plasma is in nearly perfect thermal equilibrium (the
temperature variations are only about 1 part in $10^5$),
the photons we see scattered into the line of sight
from a given fluid element have the blackbody distribution
corresponding to the temperature of that element.
Recalling that the energy density of blackbody radiation
is $\rho_\gamma\propto T^4$, we see that ${1\over4}
\delta_\gamma$ is simply the fractional temperature
variation of the fluid element.  In other words, if we
carried a thermometer to the photosphere and measured its
reading in direction $-\vec n$, it would read a fraction
${1\over4}\delta_\gamma$ different from the mean.  Now the
energy density is defined in the fluid rest frame, while
the fluid is moving with 3-velocity $\vec v_e$, so the
temperature measured by a stationary thermometer (with fixed
$x^i$) is changed by a Doppler correction $\vec v_e\cdot\vec n$.
(If the photosphere is approaching us, $\vec v_e\parallel
\vec n$, the radiation is blue-shifted.)

Comparing the CMB with a star, when we look at the surface
of a star we see the temperature of the emitting gas (to
the extent that local thermodynamic equilibrium applies),
corresponding to the ${1\over4}\delta_\gamma$ term.  For
ordinary stars the Doppler boosting and gravitational
redshift effects are negligible, although they are appreciable
for supernovae and white dwarfs, respectively.  For the
CMB, on the other hand, all four emission terms in equation
(\ref{cmblos}) are comparable in importance.

It is instructive to approximate the photosphere as an
infinitely thin layer by adopting the instantaneous
recombination approximation, according to which the
free electron fraction and hence opacity drops suddenly
at $\tau_{\rm rec}$, the conformal time of recombination
at $z\approx1100$:
\begin{equation}
  \label{instrec}
  \tau_{\rm T}(\chi)=\cases{\infty,\ \chi>\chi_{\rm rec}=
    \tau_0-\tau_{\rm rec}\ ,\cr
    0,\ \ \,\chi<\chi_{\rm rec}\ .\cr}
\end{equation}
Substituting into equation (\ref{cmblos}) and integrating
by parts the gravitational redshift term gives the result
first obtained in another form by Sachs and Wolfe in 1967
[\cite{sw}]:
\begin{equation}
  \label{cmbsw}
  \Delta(\vec n\,)\approx\left({1\over4}\delta_\gamma+\phi
  +\partial_\chi W_\gamma\right)_{\rm rec}
    +\int_0^{\chi_{\rm rec}}d\chi\,
    (2\partial_\tau\phi)_{\rm ret}
\end{equation}
where we have dropped the argument $\tau_0$ from $\Delta$,
$\chi_{\rm rec}=\tau_0-\tau_{\rm rec}$, and $W_\gamma$ is
the velocity potential for the photons ($W_\gamma=W_r$ in
the two-fluid approximation discussed above).  We could have
used the last of equations (\ref{linfluid}) to write the
sum of the intrinsic (${1\over4}\delta_\gamma$) and
gravitational ($\phi$) contributions as $\partial_\tau W_r$.
The combination of derivatives $(\partial_\tau+
\partial_\chi)W_r$ is the time derivative along the past
light cone; the results are evaluated at recombination.

While it is interesting that the primary anisotropy depends
on the velocity potential derivative along the line of sight,
there appears to be no fundamental significance to the
simple dependence in equation (\ref{cmbsw}), aside from
the fact that CMB anisotropy is produced by departures from
hydrostatic equilibrium.  (In hydrostatic equilibrium,
$\vec v_r=0$.)  If the radiation gas were in hydrostatic
equilibrium in a static gravitational potential, there would
be no primary CMB anisotropy.  Indeed, it can be shown
from purely thermodynamic arguments that $T(\vec x\,)\propto
\exp[-\phi(\vec x\,)]$ (hence ${1\over4}\delta_\gamma+\phi=
\hbox{constant}$) for a photon gas in equilibrium.

Sachs and Wolfe [\cite{sw}] showed that for adiabatic
perturbations in the matter-dominated era, on scales much
larger than the acoustic horizon ($\omega\tau\ll1$ in eq.
\ref{phietagrow}), the sum of the photospheric terms in
equation (\ref{cmbsw}) (the terms evaluated at recombination)
is ${1\over3}\phi$.  Thus, on angular scales much larger
than $1^\circ$ (roughly the size of the acoustic horizon),
the CMB anisotropy is a direct measure of the gravitational
potential on the photosphere at recombination.

\subsection{Angular Power Spectrum}

The angular power spectrum gives the mean squared amplitude
of the CMB anisotropy per spherical harmonic component.
Thus, we expand the anisotropy in spherical harmonic
functions of the photon direction:
\begin{equation}
  \label{alm}
  \Delta(\vec n\,)=\sum_{l,m}a_{lm}Y_{lm}(\vec n\,)\ .
\end{equation}
Observations of our universe give definite numbers for
the $a_{lm}$ (with experimental error bars, of course).
Theoretically, however, we can only predict the probability
distribution of the $a_{lm}$.  For statistically isotropic
fluctuations (i.e. having no preferred direction a priori),
the $a_{lm}$ are random variables with covariance
\begin{equation}
  \label{covalm}
  \left\langle a_{lm}a^\ast_{l'm'}\right\rangle=
    C_l\delta_{ll'}\delta_{mm'}
\end{equation}
where the Kronecker delta's make the $a_{lm}$'s
uncorrelated ($\delta_{ll'}=1$ if $l=l'$ and 0 otherwise).
The variance of each harmonic is given by the angular
power spectrum $C_l$; rotational symmetry ensures that,
theoretically, it is independent of $m$.

To calculate the angular power spectrum, we expand
$\phi(\vec x,\tau)$ (and the other variables) in plane
waves (assuming that the background is flat, $K=0$),
\begin{equation}
  \label{fourierexp}
  \phi(\vec x,\tau)=\int d^3k\,e^{-i\mu k\chi}
    \phi(\vec k,\tau)
    =\int d^3k\,\sum_{l=0}^\infty (-i)^l(2l+1)
    j_l(k\chi)P_l(\mu)\phi(\vec k,\tau)
\end{equation}
where $\mu=-\vec k\cdot\vec n/k$ with a minus sign
because $-\vec n$ is the radial unit vector at the
origin.  Note that many cosmologists insert a factor
$(2\pi)^{-3}$ in the Fourier integral going from
$k$-space to $x$-space; take heed of the conventions
when they matter.

In equation (\ref{fourierexp}) we have used the spherical
wave expansion of a plane wave in terms of spherical Bessel
functions $j_l(x)$ and Legendre polynomials $P_l(x)$.
This gives
\begin{equation}
  \label{delexp}
  \Delta(\vec n\,)=\int d^3k\,\sum_{l=0}^\infty
    (-i)^l(2l+1)P_l(\mu)\Delta_l(\vec k\,)
\end{equation}
where, in the instantaneous recombination approximation,
\begin{eqnarray}
  \label{deltal}
  \Delta_l(\vec k\,)&=&\left[{1\over4}\delta_\gamma(\vec k,
    \tau_{\rm rec})+\phi(\vec k,\tau_{\rm rec})\right]
    j_l(k\chi_{\rm rec})+kW_r(\vec k,\tau_{\rm rec})
    j'_l(k\chi_{\rm rec})\nonumber \\
    &&+\int_0^{\chi_{\rm rec}}d\chi\,
    2\partial_\tau\phi(\vec k,\tau_0-\chi)j_l(k\chi)\ .
\end{eqnarray}
The temperature expansion coefficient takes a simple form
in terms of $\Delta_l$:
\begin{equation}
  \label{almk}
  a_{lm}=(-i)^l4\pi\int d^3k\,Y^\ast_{lm}(\hat k)\Delta_l
    (\vec k\,)
\end{equation}
where $\hat k$ is a unit vector in the direction of $\vec k$.

To get the angular power spectrum now we must relate
$\Delta_l(\vec k\,)$ to the initial random field of potential
or entropy fluctuations that induced the CMB anisotropy.
Let us assume we have adiabatic fluctuations, for which
$\phi(\vec k,\tau\to0)=A(\vec k\,)$.  We define the CMB
transfer function
\begin{equation}
  \label{dldef}
  D_l(k)\equiv{\Delta_l(\vec k\,)\over A(\vec k\,)}\ ,
\end{equation}
which depends only on the magnitude of $\vec k$ because
the equations of motion have no preferred direction.  To
compute ensemble averages of products, we will need the
two-point function for a statistically homogeneous and
isotropic random field, whose variance defines the
power spectrum:
\begin{equation}
  \label{pspect}
  \left\langle A(\vec k\,)A^\ast(\vec k^{\,\prime})\right
    \rangle=P_A(k)\delta_{\rm D}^3(\vec k-\vec k^{\,\prime})\ .
\end{equation}
Note that many cosmologists use an alternative definition
of the power spectrum with a factor $(2\pi)^3$ inserted on
the right-hand side.  The power spectrum so defined is
greater by a factor $(2\pi)^3$ than the one given here.
[The extra factors of $(2\pi)^3$ in eq. \ref{fourierexp}
are then important.]  The definition presented here is
conventional in field theory and it ensures that $P_A(k)d^3k$
is the contribution per unit volume in $k$-space to the
variance of $A(\vec x\,)$.  Because of this interpretation,
the power spectrum is also known as the spectral density.
See [\cite{pea}] for further discussion.

For scale-invariant fluctuations (equal power on all scales,
as predicted by the simplest inflationary models), $P_A
\propto k^{-3}$.

Combining equations (\ref{covalm}), (\ref{almk}), and
(\ref{dldef}) gives the formal expression for the angular
power spectrum as an integral over the three-dimensional
power spectrum:
\begin{equation}
  \label{clint}
  C_l=4\pi\int d^3k\,P_A(k)D_l^2(k)\ .
\end{equation}
It is difficult to get a simple approximation to $D_l(k)$
that makes this analytically tractable, except in the
limit of large angular scales where the intrinsic and
gravitational contributions to $\Delta(\vec n\,)$
dominate, with $\Delta\approx{1\over3}\phi$.  Then
$D_l={1\over3}j_l(k\chi_{\rm rec})$ and the integral
may be evaluated analytically for $P_A(k)\propto k^{n-4}$
with fixed $n$.  When $n=1$ (the scale-invariant spectrum),
the result gives $l(l+1)C_l=\hbox{constant}$ or equal power
on all angular scales.  The phenomenon of acoustic peaks
in $l(l+1)C_l$ is due to the acoustic oscillations in
$\phi$ and $W_\gamma$ which modify the potentials at
recombination from their scale-invariant primeval forms.

This presentation has been based on the traditional
Fourier space representation of the potential and other
fields.  The physical interpretation of the CMB
two-point function is somewhat clearer in real (position)
space, although a more detailed analysis based on
Green's functions is needed before one can reap the
rewards.  For an introduction to the position-space
perspective see [\cite{bb00}].

\section{Structure Formation in $\Lambda$CDM Models}

This section summarizes briefly the steps needed to go from
fluctuation evolution in the early universe to predictions
of galaxy formation and large scale structure.  The
presentation will focus on techniques rather than model
results.  However, in order to simplify the discussion,
we will focus on the favored cold dark matter family
of models (with curvature and vacuum energy as options)
with adiabatic density fluctuations from inflation.

\subsection{CDM Linear Transfer Function}

During the radiation-dominated era, CDM density fluctuations
grow only logarithmically.  Once the universe becomes
matter-dominated at $\tau\sim\tau_e$ (eq. \ref{ytau}),
equation (\ref{phidd}) shows that the potential becomes frozen
in as long as vacuum energy (or other sources of pressure)
and curvature are dynamically unimportant.  Thus, from
equation (\ref{poisson}), the CDM density fluctuation
amplitude grows linearly with $a$ since $a_{\rm eq}$:
\begin{equation}
  \label{deltacphi}
  \delta_m(k,\tau>\tau_e)={-k^2\phi\over4\pi Ga^2
    \bar\rho_m}=-{2\over3}{a\over a_{\rm eq}}(k\tau_e)^2
    \phi(k,\tau_e)\ .
\end{equation}
We have used subscript ``m'' to include all cold matter
including baryons, which cluster with the CDM after
recombination.  We have also included the velocity
potential term $3(\dot a/a)W$ in equation (\ref{poisson})
as part of $\delta_m$ as it is in Bardeen's gauge-invariant
density perturbation variable $\epsilon$ [\cite{bard}].
(The velocity potential term is anyway negligible on scales
much smaller than the Hubble distance $H^{-1}$.)

To fully specify the CDM density fluctuation spectrum
we now need to know the gravitational potential, or
equivalently the density fluctuations, at the end of
the radiation era.  There is a crucial physical scale,
the Jeans length for the radiation gas at $\tau=\tau_e$,
\begin{equation}
  \label{kradj}
  k_e^{-1}={\tau_e\over\sqrt{3}}={11\,\hbox{Mpc}\over
    \Omega_mh^2}\ .
\end{equation}
(Including the effects of baryons decreases this by
up to 15\% from what is given here.)
Perturbations of wavelength longer than $2\pi/k_e$ have
always been Jeans unstable, so that $\phi(\vec k,\tau_e)
\approx A(\vec k\,)$ for $k\ll k_e$.
Note that only the acoustic horizon---not the causal
horizon or Hubble scale---enters as a length scale in
the CDM transfer function.

To get the amplitude for shorter-wavelength perturbations
which have suffered acoustic damping (i.e. they were
Jeans stable for a period of time), it is not enough
to use the radiation-era solution for the gravitational
potential in equation (\ref{deltacphi}), because the
CDM density fluctuations dominate over those of radiation
before the universe becomes matter-dominated.  Instead, we
turn to equations (\ref{poisson}) and (\ref{phietagrow}) for
$\omega\tau\gg1$ in the radiation era.  For the adiabatic
mode this gives
\begin{eqnarray}
  \label{delsubhor}
  \delta_c&=&{3\over4}\delta_r-\eta\approx{9\over2}A(\vec k\,)
    \cos(\omega\tau)-\eta\ ,\quad\hbox{where}\ \omega
      \equiv k/\sqrt{3}\nonumber \\
   &\approx&-9A(\vec k\,)\left[\ln(\omega\tau)+
    {\cal C}-{1\over2}-{\rm Ci}(\omega\tau)\right],
    \quad\tau<\tau_e\ \hbox{and}\ \omega\tau\gg1.
\end{eqnarray}
The gravitational impulse created by the acoustic oscillations
of the radiation fluid causes the CDM to oscillate slightly
but as $\omega\tau$ grows the cosine integral quickly dies and
the CDM grows logarithmically.  This growth is important for
structure formation as it allows the CDM perturbations to
dominate over the radiation perturbations enough for galaxies
to form in less than a factor 1000 expansion since recombination
when the large-scale radiation fluctuations were only a few
parts in $10^5$.

Combining results, we obtain an approximate solution for
the CDM transfer function $T_m(k)$ giving the amplitude of
CDM density fluctuations relative to the primeval potential
$A(\vec k\,)$:
\begin{equation}
  \label{tmlim}
  T_m(k)\approx\cases{-2(k/k_e)^2\ ,\quad\quad
    \quad\quad k\ll k_e\ ,\cr
    -9[\ln(k/k_e)+{\cal C}]\ ,\quad\ k\gg k_e\ ,}
\end{equation}
where
\begin{equation}
  \label{tmdef}
  \delta_m(k,\tau>\tau_e)\equiv{a(\tau)\over a_{\rm eq}}
    A(\vec k\,)T_m(k)\ .
\end{equation}
Bardeen et al. [\cite{bbks}] give a fitting formula for
$T_m(k)/k^2$, which, when normalized to 1 as $k\to0$, is
conventionally called the CDM transfer function.  However,
given the measurements of the gravitational potential made
possible from the CMB, it is preferable to retain the
normalization of equation (\ref{tmlim}) with the potential
$A(\vec k\,)$ factored out in equation (\ref{tmdef}).

The CDM linear power spectrum follows by analogy with equation
(\ref{pspect}) for the potential.  The result, for a
matter-dominated universe, is
\begin{equation}
  \label{pcdm}
  P_m(k,\tau)=\left[a(\tau)\over a_{\rm eq}\right]^2T_m^2(k)
    P_A(k)\ .
\end{equation}
The constant-curvature scale-invariant spectrum of potential
fluctuations $P_A\propto k^{-3}$ was first proposed by
Harrison [\cite{harr70}] for its naturalness properties
long before inflationary cosmology gave it a concrete basis.
With this spectrum, the CDM density spectrum has the
well-known limiting behavior
\begin{equation}
  \label{pcdmlim}
  P_m(k)\propto\cases{k\ ,\quad\quad\quad\quad
    \quad\quad\ \, k\ll k_e\ ,\cr
    k^{-3}[\ln(k/k_e)]^2\ ,\quad k\gg k_e\ .}
\end{equation}

The power spectrum is useful for calculating the variance
of the smoothed density field.  If we convolve $\delta(
\vec x,\tau)$ with a spherical window function whose
Fourier transform is $W(k)$ normalized so that $W(0)=1$,
then
\begin{equation}
  \label{denvar}
  \left\langle\bar\delta^2\right\rangle=\int d^3k\,
    P(k)W^2(k)\ .
\end{equation}
From this, we see that the CDM density fluctuations
have logarithmically divergent power at small scales.
Thus, structure formation in a CDM model proceeds by
hierarchical or bottom-up clustering with small scales
collapsing by gravitational instability before being
incorporated into larger objects.  However, the slowness
of the logarithm means that the structure formation
proceeds rapidly from small to large scales.

\begin{figure}
  \centerline{\epsfig{file=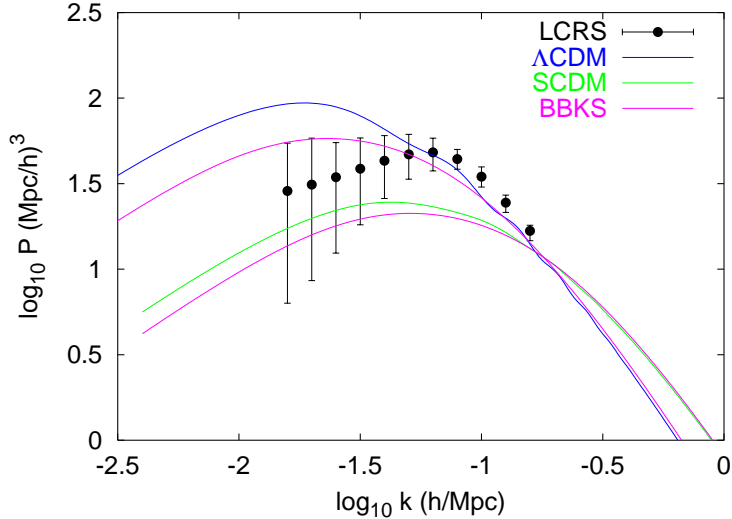,width=4.0in}}
  \caption{Theoretical CDM power spectra compared with
  the deconvolved power spectrum of galaxies from the
  Las Campanas Redshift Survey [\cite{lcrs}].  The error
  bars are strongly correlated at large scales.}
  \label{fig:pk}
\end{figure}

Figure \ref{fig:pk} compares the linear power spectra
extrapolated to $\tau=\tau_0$ (i.e. redshift 0) for
several flat CDM models.  The curves are all normalized
so that the density smoothed with a 8 $h^{-1}$ Mpc
spherical ``tophat'' has standard deviation $\sigma_8=0.9$
in agreement with observed galaxy clustering for a
``linear bias factor'' $b=\sigma_8^{-1}=1.1$.
The $\Lambda$CDM model with $(\Omega_b=0.05,\Omega_c=
0.30,\Omega_\Lambda=0.65,h=0.65)$ has the most power on
large scales, possibly too much (although an acceptable
fit can be found with slightly higher $\Omega_c$).
The SCDM model with $(\Omega_b=0.05,\Omega_c=0.95,h=0.50)$
has too little power on large scales.  When combined
with the CMB anisotropy, this provides a strong argument
for dark energy (e.g. a cosmological constant)
[\cite{tzh}].

The curves labeled ``BBKS'' show the linear power
spectra for the $\Lambda$CDM and SCDM models in the
limit $\Omega_b=0$ (with $\Omega_c=0.35$ and 1.0
for the two models, respectively).  They are obtained
from the fitting formula given by equation (G3) of
[\cite{bbks}] and agree well with full calculations
using the COSMICS code [\cite{cosmics}], which was
used for the other curves in Figure \ref{fig:pk}.
Baryons reduce the amount of small-scale power relative
to scales $k<k_e$ because they cannot experience
gravitational instability on scales smaller than the
photon-baryon Jeans length until after recombination.
The oscillations in the $\Lambda$CDM power spectrum
are also due to baryons and are similar to the
cosine integral contributions to $\delta_c(k)$ in
equation (\ref{delsubhor}).  They arise from the
acoustic oscillations of the baryons between $\tau_e$
and $\tau_{\rm rec}$.  When the power spectra are
normalized at small scales (as they effectively
are with $\sigma_8$), increasing the baryon
fraction increases the amount of large-scale power.

Aside from the baryon effect, the main parameter
affecting the CDM spectrum (and the only one in the
BBKS formula) is $k_e$.  As may be checked using equation
(\ref{kradj}), the peak of the CDM power spectrum is
closely equal to $k_e$.  Measurements of galaxy
clustering actually constrain $k_e/h=0.1\Gamma$
Mpc$^{-1}$ where $\Gamma\equiv\Omega_m h$ is a commonly
used parameter.  However, as Figure \ref{fig:pk} shows,
a second parameter, the baryon fraction $\Omega_b/
\Omega_m$, also significantly affects the CDM spectrum.

\subsection{Evolution of Large Scale Structure}

The results given above concerned the shape of the
CDM power spectrum, which is frozen in after the universe
becomes matter-dominated.  However, the amplitude grows
as $a(\tau)$ as assumed in equation (\ref{tmdef}) only
if $\Omega_m=1$.  While this is a good approximation
soon after recombination, it breaks down at small redshift
when $\Omega_m<1$ today.  In this case we must include
dark energy, curvature, and any other important contributor
to the expansion rate of the universe.

The evolution of CDM density fluctuations for $\tau\gg
\tau_e$ is given by equations (\ref{poisson}) and
(\ref{linfluid}) ignoring the radiation terms.  On
subhorizon scales for $\phi^2\ll\delta_m^2\ll1$ we have
\begin{equation}
  \label{delcdd}
  \partial_\tau^2\delta_m+{\dot a\over a}\partial_\tau
    \delta_m=4\pi Ga^2\bar\rho_m\delta_m={3\over2}
    \Omega_mH_0^2\,{\delta_m\over a}\ .
\end{equation}
Because the coefficients are spatially homogeneous, the
time and space dependences of the solution factor:
$\delta_m(\vec x,\tau)=\delta_+(\vec x\,)D_+(\tau)+
\delta_-(\vec x\,)D_-(\tau)$.
For $\Omega_m>0$, $D_+$ grows with time and $D_-$
decays.  For $\Omega_m=1$ the growing solution is
$D_+=a$.  For $\Omega_m<1$ the solutions depend on the
curvature $K$ (sometimes expressed as $\Omega_K=-K/H_0^2$)
and on the equation of state of the dark energy.  See
Peebles [\cite{p80}] and Peacock [\cite{pea}] for
examples.

Peculiar velocities offer another probe of large-scale
structure that is sensitive directly to all matter,
dark or luminous.  The divergence of peculiar velocity
$\vec v=d\vec x/d\tau$ depends on the growth rate of
density fluctuations:
\begin{equation}
  \label{thetam}
  \theta_m\equiv\vec\nabla\cdot\vec v_m=-\partial_\tau
    \delta_m=-\left(d\ln D_+\over d\ln a\right)aH
    \delta_m\quad\hbox{for}\ \delta_m^2\ll1\ .
\end{equation}
Measurements of galaxy velocities actually give
$\vec v/aH$, so that the ratio of velocity divergence
to density fluctuation gives the logarithmic growth
rate of density fluctuations.  An empirical fit to
a range of models gives $d\ln D_+/d\ln a=f(\Omega_m)
\approx\Omega^{0.6}$.  In practice, the galaxy number
density may be a biased tracer of the total matter
density, conventionally parameterized by the bias
factor $b$: $\delta_{\rm galaxies}=b\delta_m$.
Comparing galaxy clustering and velocities then allows
determination of $\Omega_m^{0.6}/b$.

\subsection{Numerical Simulation Methods}

Linear theory breaks down once dark matter clumps
become significantly denser than the cosmic mean.
Once this happens, full numerical simulations are
needed to accurately follow the nonlinear dynamics
of structure formation.  Moreover, radiative processes,
star formation, and supernova feedback have a strong
effect on the baryonic component once galaxies begin
forming.  Although these processes can only be included
in simulations in a phenomenological way, numerical
simulations are still the best way to advance our theoretical
understanding of structure formation deep into the
nonlinear regime.

Here we present only a brief synopsis of numerical simulation
methods.  For a comprehensive review of both the methods
and their applications see [\cite{eb98}].

Particle methods are used for representing the dark matter.
The phase space of dark matter is represented by a sample
of particles evolving under their mutual gravitational
field:
\begin{equation}
  \label{nbody}
  {d\vec x_i\over d\tau}=\vec v_i\ ,\quad
  {d\vec v_i\over d\tau}=-{\dot a\over a}\vec v_i-
    \vec\nabla\phi\ ,\quad
  \nabla^2\phi=4\pi Ga^2\left[\sum_jm_j\delta_{\rm D}
    (\vec x-\vec x_i)-\bar\rho_m(\tau)\right]
\end{equation}
where $\delta_{\rm D}$ is the Dirac delta function.
In practice, the particles are spread over a softening
distance to avoid unphysical two-body relaxation due to
large-angle scattering.  (A simulation may use particles
of mass $10^9\ M_\odot$, leading to unphysical tight
binaries if the forces are not softened.)  The art of
dark matter simulation is mainly in calculating forces
much faster than $O(N^2)$ for a softening distance
several orders of magnitude smaller than the simulation
volume.  A variety of algorithms and codes are available;
see [\cite{eb98}] for details.

The initial conditions for dark matter positions and
velocities are given from linear theory by the
Zel'dovich approximation [\cite{za}], which relates
the comoving position to its initial value $\vec q$,
a Lagrangian variable (i.e. one that labels each
particle):
\begin{equation}
  \label{zax}
  \vec x(\vec q,\tau)=\vec q+D_+(\tau)\vec\psi(\vec q\,)
    \ ,\quad \vec\nabla_q\cdot\vec\psi=-\delta_m\ .
\end{equation}
The initial density fluctuation field $\delta_m(\vec x,
\tau)$ is obtained as a sample of a gaussian random field
with appropriate power spectrum.

Baryon physics is added by numerical integration of the
fluid equations (\ref{conteq}) and (\ref{eulereq}) with
the addition of heating and cooling, ionization, etc.
There are two classes of methods in widespread use in
cosmology: smoothed-particle hydrodynamics (SPH) and
grid-based methods.  SPH is an extension of the N-body
dynamics of equations (\ref{nbody}) to include pressure
forces and thermodynamics.  It has the advantage of
concentrating the spatial resolution and computational
effort where the baryons are densest, with the drawback
of relatively large numerical viscosity and diffusivity
owing to the particle discreteness.  Grid-based methods
use finite-difference techniques to approximate the
fluid equations on a (usually) regular lattice.  High
accuracy methods developed for aerodynamic and other
applications provide excellent resolution of shock waves,
but with a resolution limited by the grid.  Adaptive
mesh refinement techniques are now being used to refine
the resolution where needed, resulting in the ability
to follow structure formation to much higher densities
than previously possible [\cite{abn}].

\begin{figure}
 \centerline{\epsfig{file=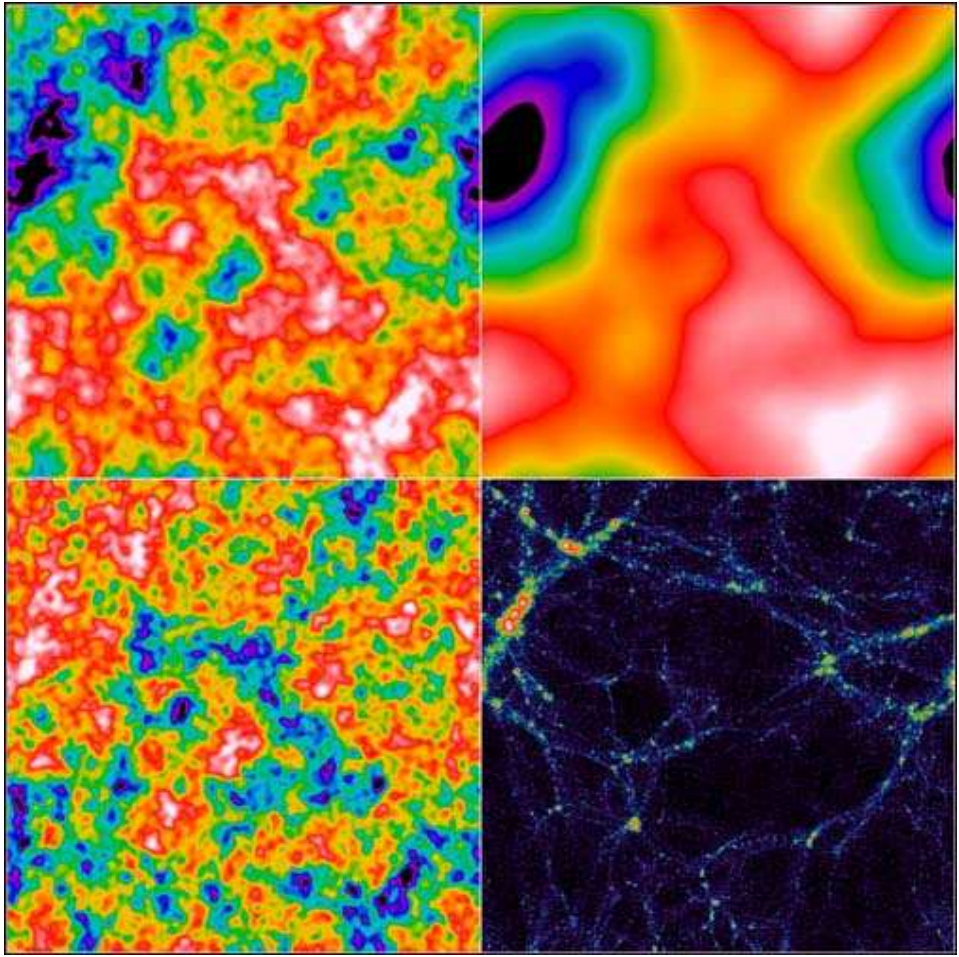,width=5.0in}}
  \caption{Potential and density for a mixed cold+hot
    dark matter model with $\Omega_\nu=0.2$
    [\cite{mabert94}].  A slice
    of size $50 h^{-1}$ Mpc is shown.  Upper left:
    potential at the end of inflation.  Upper right:
    potential at the end of recombination.  Lower
    left: density at the end of recombination.
    Lower right: density at redshift 0.}
  \label{fig:pd}
\end{figure}

We conclude this introduction to numerical simulation
methods with a mosaic of gravitational potential and
dark matter density from a high-resolution simulation
of the mixed cold+hot dark matter model.
Figure \ref{fig:pd} shows the linear theory and nonlinear
simulation results.  On the scales shown, $\Lambda$CDM
would be almost indistinguishable.

The gravitational potential shown in Figure \ref{fig:pd}
is a statistical fractal at the end of inflation, viz.
a gaussian random field with power spectrum $P_A\propto
k^{-3}$.  During the radiation era, acoustic damping
smooths the potential on scales smaller than $k_e^{-1}$.
Because the density fluctuations are related to the
potential by two spatial derivatives, the corresponding
linear density field is dominated by high spatial
frequency components.  Indeed, if not for the smoothing
imposed by finite spatial resolution, the variance of
density fluctuations would be logarithmically divergent.
It is instructive to compare the post-recombination
density with the inflationary potential.  The density
is almost proportional to the potential but with
opposite sign.  This is because the density/potential
transfer function (\ref{tmlim}) varies only logarithmically
on small scales.

The nonlinear evolution of the density field causes the
mass to cluster strongly into wispy filaments and dense
clumps; most of the volume has low density.  The
intermittency of structure is a sign of strong deviations
from gaussian statistics.

\subsection{Press-Schechter Theory}

Numerical simulations show that most of the mass falls
into dense self-gravitating clumps.  As a crude
approximation, let us suppose that the mass profile of
every clump is spherically symmetric.  In this case
every gravitationally bound shell will eventually fall
into the center of the clump, at a time that is dependent
on the enclosed mass and energy of the shell.  For
linear growing mode initial conditions, collapse
occurs when linear theory would predict that the
mean density contrast interior to the shell is
$\delta_c=1.686$ [\cite{pea}].  Thus, in the spherical
model, the distribution of clump masses is simply dependent
on the statistics of the smoothed linear density field.

Press and Schechter [\cite{ps}] devised an elegant method
for computing the mass distribution of gravitationally
bound clumps using the simple spherical model.  Their
procedure is first to smooth the linear density field
by convolution with a spherical ``tophat'' of radius
$R$ enclosing mean mass $M(R)=(4\pi/3)\bar\rho_mR^3$:
\begin{equation}
  \label{delsmo}
  \bar\delta_m(\vec x\,)=\int d^3x'\,W_R(\vert\vec x-
    \vec x^{\,\prime}\vert)\delta_m(\vec x^{\,\prime})\ ,
  \quad W_R(r)={3\over4\pi R^3}\cases{1,\ r<R\ ,\cr
    0,\ r>R\ .}
\end{equation}
According to the Press-Schechter ansatz, the regions of
the initial density field with $\bar\delta>\delta_c=1.686$
at time $\tau$ have collapsed into objects of mass at least
$M$.  It is easy to see that this would be true for a
single spherical density perturbation; the genius is
applying this to the entire density field.

Under this model, the fraction of mass in the universe contained
in clumps more massive than $M$ is
\begin{equation}
  \label{massfrac}
  {1\over\bar\rho_m}\int_M^\infty dM\,M{dn\over dM}=
  P(\bar\delta>\delta_c\vert M,\tau)\ .
\end{equation}
On the left-hand side, $(dn/dM)dM$ is the number density
of clumps of mass in $(M,M+dM)$; the integral gives the
total mass density in clumps more massive than $M$.
Dividing by $\bar\rho$ gives the mass fraction.  On the
right-hand side, $P(\bar\delta>\delta_c\vert M,\tau)$
is the probability that a randomly chosen point in the
linear density field has $\bar\delta>\delta_c$ at time
$\tau$, where $M$ defines the smoothing scale of $\bar\delta$.
For a gaussian random field of density fluctuations, this
probability is simply
\begin{equation}
  \label{perfc}
  P={1\over2}\hbox{erfc}\left(\delta_c\over\sigma\sqrt{2}\right)
\end{equation}
where erfc is the complementary error function and $\sigma^2=
\sigma^2(M,t)=\langle\bar\delta^2\rangle$ is the variance of
the smoothed density.  From equation (\ref{massfrac}) we can
deduce $dn/dM$ by differentiating with respect to $M$.

Now, as $M\to0$, $\sigma(M)$ grows without bound for CDM
models with logarithmically diverging power on small scales,
so that $P(\bar\delta>
\delta_c\vert M=0,\tau)={1\over2}$.  Half of the mass
initially is in overdense regions, half is in underdense
regions $\bar\delta<0$ which in linear theory can never
collapse.  However, this is plainly unphysical, as simulations
show that virtually all of the mass, including that in
initially underdense regions, accretes onto dense clumps.
Thus, Press and Schechter suggested doubling $dn/dM$ so that
all of the mass resides in dense clumps.  Thus, their formula
for the mass distribution of clumps is
\begin{equation}
  \label{psdist}
  {dn\over dM}={2\bar\rho\over M}{\partial P\over\partial M}\ .
\end{equation}

Although realistic gravitational collapse is far more complex
than imagined in this derivation, equation (\ref{psdist})
gives surprisingly good agreement with clump masses measured
in cosmological N-body simulations [\cite{lc96}].  The
Press-Schechter model and its generalizations have proved
to be a very useful tool for comparing ab initio theories
(power spectrum and cosmological model) with observations
of galaxies and galaxy clusters.

\section{Conclusions}

These lecture notes have presented a few of the theoretical
elements useful for understanding the microwave background
and structure formation.  Cosmology is rapidly changing with
the advent of precision measurements of CMB anisotropy, with the
observation of galaxies at $z>3$, and with large new redshift
surveys.  Theoretical research has become more phenomenological,
with a focus on providing the framework for interpreting and
analyzing the new data.  Despite this trend---or perhaps because
of the high quality of new data---theoretical cosmology still
requires further development of the basic processes of structure
formation.  Either way, I hope that the material presented in
these lectures will be useful to cosmology graduate students.

\acknowledgments

I would like to thank Sergei Bashinsky for help with the
two-fluid solutions.  I am grateful to the organizers,
the other speakers, and the students of the Cosmology 2000
summer school for the fruitful, stimulating atmosphere in
Lisbon.  Special thanks are due to the organizers for their
wonderful hospitality.  This work was supported by NSF grant
AST-9803137.

\thebibliography

\bibitem{lesh}[1] E. Bertschinger, in Cosmology and Large Scale Structure,
  proc. Les Houches Summer School, Session LX, ed. R. Schaeffer, J. Silk,
  M. Spiro, and J. Zinn-Justin, Elsevier Science, Amsterdam (1996) 273.
\bibitem{mabert}[2] C.-P. Ma and E. Bertschinger, \apj\ 455 (1995) 7.
\bibitem{lifsh}[3] E. M. Lifshitz, J. Phys. USSR 10 (1946) 116.
\bibitem{mtw}[4] C.W. Misner, K.S. Thorne, and J.A. Wheeler, Gravitation,
  Freeman, San Francisco (1973).
\bibitem{gpb}[5] Gravity Probe B website, http://einstein.stanford.edu/.
\bibitem{bard}[6] J.M. Bardeen, \prd\ 22 (1980) 1882.
\bibitem{za}[7] Ya.B. Zel'dovich, \aap\ 5 (1970) 84.
\bibitem{pd68}[8] P.J.E. Peebles and R.H. Dicke, \apj\ 154 (1968) 891.
\bibitem{cmbfast}[9] U. Seljak and M. Zaldarriaga, \apj\ 469 (1996) 437.
\bibitem{hswz98}[10] W. Hu, U. Seljak, M. White, and M. Zaldarriaga,
  \prd\ 57 (1998) 3290.
\bibitem{gl}[11] P. Schneider, J. Ehlers, and E.E. Falco, Gravitational Lenses,
  Springer-Verlag, Berlin (1992).
\bibitem{selgl}[12] U. Seljak, \apj\ 463 (1996) 1.
\bibitem{peeyu}[13] P.J.E. Peebles and J.T. Yu, \apj\ 162 (1970) 815.
\bibitem{sw}[14] R.K. Sachs and A.M. Wolfe, \apj\ 147 (1967) 73.
\bibitem{pea}[15] J.A. Peacock, Cosmological Physics, Cambridge Univ.
  Press (1999).
\bibitem{bb00}[16] S. Bashinsky and E. Bertschinger, astro-ph/0012153.
\bibitem{bbks}[17] J.M. Bardeen, J.R. Bond, N. Kaiser, and A.S. Szalay,
  \apj\ 304 (1986) 15.
\bibitem{harr70}[18] E.R. Harrison, \prd\ 1 (1970) 2726.
\bibitem{lcrs}[19] H. Lin et al., \apj\ 471 (1996) 617.
\bibitem{tzh}[20] M. Tegmark, M. Zaldarriaga, and A.J.S. Hamilton,
  astro-ph/0008167.
\bibitem{cosmics}[21] E. Bertschinger, astro-ph/9506070 and
  http://arcturus.mit.edu/cosmics/.
\bibitem{p80}[22] P.J.E. Peebles, The Large-Scale Structure of the
  Universe, Princeton Univ. Press (1980).
\bibitem{eb98}[23] E. Bertschinger, \araa\ 36 (1998) 599.
\bibitem{abn}[24] T. Abel, G.L. Bryan, and M.L. Norman, \apj\ 540
  (2000) 39.
\bibitem{mabert94}[25] C.-P. Ma and E. Bertschinger, \apjl\ 434
  (1994) L5.
\bibitem{ps}[26] W.H. Press and P. Schechter, \apj\ 187 (1974) 425.
\bibitem{lc96}[27] S. Cole and C. Lacey, \mnras\ 281 (1996) 716.

\endthebibliography

\end{document}